\documentstyle[epsfig]{aipproc}

\begin{document}
\title{Preheating and Supergravity}

\author{Lev Kofman}
\address{Canadian Institute for Theoretical Astrophysics\\
University of Toronto, M5S 3H8 Canada}

%\lefthead{LEFT head}
%\righthead{RIGHT head}
\maketitle

\begin{abstract}
In this talk \footnote{Invited talk to be published in the Proceedings of
 the 8th Canadian Conference on Reneral Relativity and Relativistic
 Astrophysics, McGill University, June 10-12, 1999}
 recent developments of  the theory of preheating after
inflation are briefly  reviewed. 
In  inflationary cosmology, the  particles constituting the Universe are
created after inflation due to their interaction
with moving inflaton field(s) in the process of reheating.
In inflationary models motivated by supergravity, both 
bosons and  fermions are created.
 In the bosonic sector, the leading channel
of particle production is the non-perturbative regime of parametric
resonance dominated by
 those  bosons which are created exponentially fast
 with the largest characteristic exponent.
In the fermionic sector, the leading channel  corresponds to the regime
of parametric excitation of fermions, 
 which respects Pauli blocking
but  differs significantly from the perturbative expectation.
In supergravity  we also have to consider 
production of gravitinos and moduli fields, which are cosmologically
dangerous relics.  We discuss the derivation of the gravitino equations
 in curved space-time  with moving  background scalars.
 We describe recent results on the production of
 gravitinos 
from preheating, which may put strong constraints on the inflationary
models. 

\end{abstract}

\subsection*{Preheating after Inflation}
According to  the inflationary scenario,
 the Universe initially expands quasi-exponentially
in a vacuum-like state without entropy or particles.
At the stage of inflation, all energy is contained 
 in a classical slowly moving inflaton field $\phi$.
The fundamental
Lagrangian ${\cal L}(\phi, \chi, \psi, A_i, h_{ik}, ...)$
contains the inflaton part with the potential  $V(\phi)$
and other fields
which give subdominant contributions to gravity.
The Friedmann equation for the scale factor $a(t)$ and
the Klein-Gordon equation for $\phi(t)$
determine the evolution of the background fields.
In most of the models, soon after the end of inflation,
an almost homogeneous inflaton field $\phi(t)$   coherently
oscillates with a very large  amplitude of the order of the Planck mass
around the minimum of its potential. This scalar field can be considered as
a coherent superposition of inflatons  with zero momenta.
 The amplitude of  oscillations  gradually
decreases not only because of the expansion of the
universe, but also because energy is transferred to particles
created by the oscillating field.
At this stage
we shall recall the rest of the fundamental Lagrangian
  which
includes all the fields interacting with inflaton.
These interactions
  lead  to the  creation of many ultra-relativistic
particles from the inflaton.
Gradually, the inflaton field decays and transfers
all of its energy  non-adiabatically  to the created
particles.
In this scenario
 all the matter constituting the universe
is created from this process of reheating.
If the creation of particles is sufficiently slow,
   the  particles would   simultaneously
 interact with each other and come to a state of thermal equilibrium
at  the reheating temperature $T_r$.
This gradual reheating can be treated with the perturbative theory of
particle creation and thermalization.
However, typically particle production from coherently oscillating
inflatons occurs not in the perturbative regime but in the non-perturbative
regime of parametric excitation.
Indeed, let us consider a simple toy  model 
of chaotic inflation with the  quadratic
potential $V(\phi)={1 \over 2} m_{\phi} \phi^2$ and 
${\cal L}_{int} =-{1 \over 2} g^2 \phi^2 \chi^2$
 describing the interaction between the 
 inflatons and other massless
Bose particles $\chi$. 
The quantum scalar field $\hat \chi$ in a flat FRW background has
 the eigenfunctions $\chi_{k}(t)\, e^{ -i{{\bf k}}{{\bf x}}}$
 with comoving momentum ${\bf k}$.
 The temporal part of the eigenfunction   obeys the equation
\begin{equation}
\ddot \chi_k + 3{{\dot a}\over a}\dot \chi_k + {\left(
{{\bf k}^2\over a^2}   - \xi R + g^2\phi^2 \right)} \chi_k = 0  \label{2}
\end{equation}
with vacuum-like initial conditions: $ \chi_k \simeq {e^{ -ikt} \over
 \sqrt{2k}}$
in the far past. The
coupling to the curvature $\xi R$ will not be important
in the presence of the interaction (but would lead to gravitational
preheating in the absence of the interaction).
In this model, the inflaton field $\phi(t)$   coherently
oscillates as $\phi(t) \approx \Phi(t) ~
 \sin{\left( m_{\phi}t \right)}$, with the amplitude
 $ \Phi(t)= {M_p \over \sqrt{3\pi}}\cdot{1  \over m_{\phi}t}$
 decreasing as the universe expands.
The smallness of $g^2$ alone does not necessarily lead to the
perturbative excitation of $\chi_k$ modes.
To check whether the interaction term $g^2\phi^2$ in eq. (\ref{2})
is perturbative or not, it is convenient to use
a new time variable $z=mt$ and the essential
dimensionless coupling parameter $q= {{g^2 \Phi^2} \over{ m^2}}$.
 Scalar metric fluctuations in this model
are compatible with cosmology if  the  inflaton mass is 
 $m  \simeq 10^{-6}  M_p$; therefore, 
 it is expected that $q \simeq 10^{10} g^2 \gg 1$
for not negligibly small $g^2$.
In fact, a consistent setting for the problem of $\chi$-particle
creation from the $\phi$-inflaton requires $q \gg 1$
 even without additional assumptions about $g^2$
\cite{FKL1}. Indeed, it is known that
if we have two scalars $\phi$ and $\chi$, then 
the latest stage of inflation will be driven by the lightest
scalar. The square of the effective mass of the $\chi$-field includes a term
$g^2\phi^2$. Inflation is driven by the $\phi$-field if its square mass $m^2$
is smaller than $g^2\phi^2$. This leads to the condition $q \gg 1$,
i.e. to the creation of $\chi$-particles in the resonance regime.

\subsection*{Supergravity and the Early Universe}
To make the next step beyond toy models of 
 particles interactions with the inflaton,  we have to choose
the   fundamental
Lagrangian ${\cal L}(\phi, \chi, \psi, A_i, h_{ik}, ...)$.
We may expect that the low-energy physics of the  early universe
will be described by the general four-dimensional
$N=1$  supergravity-Yang-Mills-matter theory \cite{SUGRA}.
A rather lengthy $N=1$ phenomenological supergravity Lagrangian \footnote{
When I show a viewgraph with the full supergravity Lagrangian
at the cosmology conferences, somehow people start to laugh hysterically.}
begins with the terms
\begin{eqnarray}
e^{-1}{\cal L}&=&-{1 \over 2}M_P^2 R- 
\hat\partial_{\mu}  \Phi^i\hat\partial^{\mu}  \Phi_i+
 e^K\left({\cal D}^iW {\cal D}_iW  -3 {{WW^*} \over M_P^2} \right)\nonumber\\
&-&\bar \chi _j  \not\!\! {\cal D} \chi^i- \bar \chi^i
 \not\!\! {\cal D} \chi_j -\left( e^{K/2} {\cal D}^i{\cal D}^j  W\bar \chi _i\chi _j + h.c. \right)\nonumber\\
&-& {1 \over 2}\bar \psi_\mu R^\mu +\left(
{1 \over 2}e^{K/2} W\bar \psi _{\mu R} \gamma ^{\mu \nu }\psi
_{\nu R}+ \bar \psi_{\mu L}{ \not\!\hat \partial } \Phi^i
 \gamma^\mu \chi_i+\bar  \psi_R\chi _i  e^{K/2} {\cal D}^iW+ h.c.
\right)\nonumber\\
&+&\,... \label{lag}
\end{eqnarray}
A particular  choice of the form of the Lagrangian  is motivated and 
notations are given in \cite{grav}. In Eq. (\ref{lag}) we 
choose the minimal K\"{a}hler potential $K={{\Phi_i\Phi^i} \over {M_P^2}}$,
where $\Phi^i$ is the complex conjugate of $\Phi_i$. 
 The last term of the 1st
line is the scalar potential $V(\Phi_i)$. The equations of motion
 based on the first line
should describe inflation, which is a challenging problem by itself.
For simplicity we will take a prototype model of the superpotential
$W= \sqrt{\lambda}\Phi^3/3$, which for  $\vert\Phi\vert \ll
M_P$ leads to the effective potential $\lambda\Phi^4$.
We will illustrate some effects of preheating in supergravity
with this model.

\subsection*{Preheating  of Bosons}
 Let us briefly recall
the basics of the bosonic preheating.
Consider the creation of  $\chi$-particles due to the $g^2\chi^2\phi^2$
interaction.
To get a feeling for the way in which $\chi$-particles are created from
inflaton oscillations, we have to understand the character of 
solutions of eq. (\ref{2}) for the mode functions $\chi_k(t)$.
For preheating of bosons we need the first line of the Lagrangian
(\ref{lag}). Consider the  model with
 the inflaton  potential
  $V(\phi)={1 \over 4} \lambda \phi^4$
in an  expanding universe. The problem of particle
creation in this theory  can be reduced to a similar problem
in Minkowski space-time. This can be realized with
the conformal transformation of the scalar field
$\phi\to  \varphi=a \phi   $
and with the conformal time variable 
  $\tau = \sqrt{\lambda}  \tilde\varphi
\int {dt\over a(t)}$. Therefore, ${1 \over 4} \lambda \phi^4$
theory is sometimes dubbed the conformal theory.
In conformal variables, the
Klein-Gordon equation for  $\varphi(\tau)$ is reduced to an equation in
flat space-time. Its solution is 
$\varphi(\tau) \approx   \tilde\varphi  ~f(\tau)$,
where the   amplitude of the oscillations
 $\tilde\varphi$ is constant 
until the backreaction of created particles is taken into account.
The time-dependence of the
oscillations in this theory  is not sinusoidal, but given by an elliptic
 function
$f(\tau)=  cn \left( \tau, { 1 \over \sqrt{2}}\right)$.
Eq.~(\ref{2})   for quantum fluctuations $\chi_k$
 can be simplified in this theory. 
Using a conformal transformation of the mode function
$ X_k(t)= a(t)\chi_k(t) $ in
   Eq.~(\ref{2}) we obtain
\begin{equation}
X_k''  +  {\left(\kappa^2  + q f^2(\tau) 
 \right)} X_k  = 0 \ ,
\label{10}
\end{equation}
where $\kappa^2={ k^2 \over  \lambda   \tilde\varphi^2}$,
and $q= {g^2\over \lambda}$.
The equation for fluctuations does not depend on the
expansion of the universe and is completely reduced to the similar
problem in Minkowski space-time.
This is a special feature of the
conformal  theory
 ${1 \over 4} \lambda \phi^4  + {1 \over 2} g^2 \phi^2 \chi^2 $.
The mode equation  (\ref{10})
belongs to the class of   Lam\'{e} equations.
The combination of parameters  $q=g^2/\lambda$   ultimately
defines the structure of the parametric resonance in this theory.
This means that the condition of  a  broad parametric resonance
does not require a large initial amplitude of the inflaton field,
 as for the quadratic potential.
The strength of the resonance
depends non-monotonically on the value of the $q$ parameter. 
The stability/instability chart of the  Lam\'{e} equation (\ref{10})
in the variables
$\left(\kappa^2, {{g^2} \over \lambda} \right)$ was constructed
in \cite{GKLS}.
To see how the general theory  works,
let us consider  
parametric resonance of inflaton fluctuations fluctuations $\phi_k$ due to
the self-interaction of inflaton  field in $\lambda \phi^4$ theory.
Using  conformal transformation of the mode functions
$ \varphi_{k}(\tau)= a(t)\phi_{k}(t) $,
the equation  for  $\varphi_k$ 
 can be 
reduced 
to the general equation  (\ref{10}) with the particular value 
$q=3$. 
 The equation for fluctuations $X_k$ when $q=3$ can be
 solved analytically \cite{GKLS}.
The resonance in the ``inflaton'' direction
$\phi$ is weak; the maximal value of the characteristic exponent 
of the fluctuations $\phi_{k} \propto e^{\mu_1 \tau}$  is
 $\mu \approx 0.036$.
Let us however consider a supersymmetric version of the conformal theory.
In the first line of the Lagrangian (\ref{lag})
we will put  the superpotential $W= \sqrt{\lambda}\Phi^3/3$,
which at $\phi \ll M_P$ gives us
a scalar potential with the conformal properties 
$V({\bf }\Phi)={1 \over 4} ({\bf }\Phi^2)^2$.
In supersymmetric theories, all scalar are complex.
The scalar field ${\bf }\Phi$ will have two components $(\phi, \bar \phi)$.
Let us assume that inflaton direction corresponds to the real
component $\phi$, Re~$\Phi={\phi \over \sqrt{2}}$.
 and initially $\bar \phi=0$, Im~$\Phi=0$.
The equation for the mode function of fluctuations
in the  direction $\bar \phi$ can be
obtained with the 
conformal transformation 
 $ \bar \varphi_{k}(\tau)= a(t)\bar \phi_{k}(t) $ and can be reduced 
reduces 
to the general equation  (\ref{10}) with  the parameter
$q=1$. Again, in this case the problem can be solved analytically
\cite{GKLS}. The factor $q=1$ instead of $3$ in  eq.~(\ref{10})
makes a big difference, which manifests the subtlety of the parametric resonance.
 The resonance in the   direction $\bar \phi$
is much   stronger
and broader than the resonance in the inflaton direction,
$\bar \phi_{k} \propto e^{\bar \mu \tau}$, with
$\bar \mu \approx 0.147$.
Thus, the supergravity generalization of bosonic preheating in the
 conformal theory makes a big difference 
in the bosonic preheating due to the self-interaction. 
Note that the character and strength of the parametric resonance for the general
equation (\ref{10}) depends on the shape of the effective inflaton potential
 $V(\phi)$. Moreover, 
the investigation of the
resonance in an expanding universe typically cannot  be reduced to the 
study of the regular stability/instability chart. If theory is not conformal, 
(say due to the mass term $m^2 \phi^2$ in $V(\phi)$)
the parameter $q$  in an expanding universe  is  time-dependent.
 For the broad resonance case $q \gg 1$
this parameter can jump over a number of instability bands within a
single oscillation of the inflaton field,
and the concept of stability/instability bands is
inapplicable here. 
Parametric resonance in this case is a stochastic process \cite{KLS}.

\subsection*{Parametric Excitations of Fermions}
A simple model for the inflaton's interaction with Fermi particles $\chi$ is 
 a Yukawa term 
$ h \bar \chi \phi \chi$. For instance, consider second line of eq.~(\ref{lag})
and ignore mixing between $\chi$ and $\psi_{\mu}$ (which corresponds to the
rigid SUSY limit).
For our toy model with $W={1 \over 3}\sqrt{\lambda} \Phi^3$ 
and $\phi \ll M_P$, the  ``mass'' term of the
chiral fermion $\chi$ is equal to $\sqrt{2\lambda}\phi$,
 which corresponds to the
inflaton-fermion interaction $ \sqrt{2\lambda} \bar \chi \phi \chi$.
For fermions, the Pauli exclusion principle prohibits the 
occupation number from exceeding $1$.  For this reason,
it has  been silently assumed that  fermions are created  in the three-legs
perturbative process  $\phi \to  \bar\chi\chi$
 where individual inflatons  decay independently into 
 pairs of  $\psi$-particles.
Let us, however, consider the Dirac equation for a
massless quantum Fermi field $\chi(t, \vec x)$:
\begin{equation}
\left[  \gamma^{\mu} \nabla_{\mu} +h \phi(t)\right] \chi=0 \ ,
\label{5}
\end{equation}
where $ \nabla_{\mu}$ is the derivative with the spin connection.
We are using the representation of gamma matrices where
 $\gamma_0=diag(i, i, -i, -i)$.
Here, similar to the bosonic case, the  inflatons producing fermions  also 
 act not as individual particles  but as a coherently oscillating
field $\phi(t)$. 
Let us consider more general  model 
 ${1 \over 4}\lambda \phi^4 + h\bar \psi \phi \psi$.
This is a conformal theory in the sense that
the problem of fermion production by the inflaton $\phi$
in an expanding universe can be 
  reduced to equations  in Minkowski space-time.
Indeed, let us perform a conformal transformation of the involved
 fields, $\varphi \equiv a \phi$ and $\Psi \equiv a^{3/2}\chi$, 
and use a conformal  time variable,
 $\tau$ as in the previous section.
The equation  for the eigenfunctions of the 
quantum fluctuations in this theory 
 can be reduced to a second-order equation for an
auxillary field $X(\tau, \vec x)$,
so that
 $\Psi= \left[  \gamma^{\mu} \nabla_{\mu} +h \varphi\right]
 X$. The eigenmodes of the auxillary field have the form
$ X_k(\tau)e^{ +i{{\bf k}} \cdot {{\bf x}}}R_r$, 
where the $R_r$ are  eigenvectors
of the Dirac matrix $\gamma^0$ with eigenvalue $+1$.
The temporal part of the eigenmode
obeys an oscillator-like equation with a complex frequency
which depends periodically on time
\begin{equation}
 X_k''  +  {\left(\kappa^2  +
 q f^2 -i   \sqrt{q}  f' 
 \right)} X_k  = 0 \ .
\label{100}
\end{equation}
The dimensional comoving momentum $k$ enters the equation in the combination
 ${ k^2 \over  \lambda   \tilde\varphi^2} \equiv \kappa^2$;
therefore, the natural units of momentum are $\sqrt{\lambda} \tilde\varphi$.
The background oscillations enter in the form
 $f(\tau)$ of the previous section,  having
unit amplitude.
The imaginary part of the frequency in  Eq.~(\ref{100}) guarantees
the Pauli blocking for the occupation number $n_k$.
The results for  $n_k$ can be formulated as
follows \cite{GK}. Even though the Yukawa interaction  contains a small
factor $h$, one cannot use the perturbation
expansion in ${h}$. This is because the frequency of the
background field oscillations is  proportional to another small
parameter
$\sqrt{\lambda}$.
The combination of the coupling parameters ${h^2 \over \lambda} \equiv q$
ultimately determines the strength of the effect.
The growth of fermionic modes  occurs
in the non-perturbative regime of parametric excitation.
 The modes
get fully excited with occupation numbers $n_k \simeq 1$ 
 within
 tens of oscillations of the field $\phi$, and the width of the
parametric excitation of fermions in momentum space is about
$q^{1/4} \sqrt{\lambda}\phi_0$.
 For instance, in the case of $h=\sqrt{2\lambda}$,
$q=2$, the modes will be excited in about ten oscillations,
and the width will be about $\sqrt\lambda\phi_0$.

\subsection*{Equations for the Gravitino}
The rest of my talk is based on the recent project \cite{grav}\footnote{
All credit  for the correct   results below go to \cite{grav}.
All incorrect deviations is my fault.}.
Let us now consider the third line in the supergravity Lagrangian (\ref{lag}),
which describes the gravitino field $\psi_{\mu}$.
In a general background metric and in the presence of complex scalar
fields with non-vanishing VEV's,  the  equation for the
gravitino  has on the left hand side  the kinetic part $R^{\mu} 
\equiv \epsilon^{\mu\nu\rho\sigma}\gamma_5\gamma_{\nu}
{\cal D}_{\rho}\psi_{\sigma}$, 
 and a rather lengthy right hand
side. We will use the long
derivative ${\cal D}_{\mu}$ with the spin connection and Christoffel
symbols, for which ${\cal D} _\mu \gamma_\nu=0$.
Apart of varying
gravitino mass $m$, the right hand side
contains a chiral
 connection  and various mixing terms
like those in the 3rd  line of (\ref{lag}).
 For a self-consistent setting of the
problem, the gravitino equation should be supplemented by the
equations for the fields mixing with gravitino, $\chi_i$ from
(\ref{lag}),
 as well as by the
equations determining the gravitational background and the
evolution of the scalar fields.
Let us make some simplifications. We consider the supergravity
multiplet and a single chiral multiplet
 containing
a complex scalar field $\Phi$ with   a single chiral fermion $\chi$.
This is a simple non-trivial extension which allows us to study the
gravitino with a non-trivial FRW  cosmological metric
 supported by the scalar field.
 A nice feature of this model is that
   the chiral fermion $\chi$
can be gauged to zero so that the mixing between $\psi_{\mu}$ and
$\chi$ in (\ref{lag}) is absent. We also can choose the
non-vanishing VEV of the scalar field (inflaton) in the real direction,
as in the previous sections.
First we will derive the equation for a spin 3/2 field in a
curved  background metric with non-vanishing VEVs for the scalar fields.

\subsubsection{Spin 3/2 Field Equations in  External gravitational
and Scalar Fields}
From (\ref{lag})  we can obtain  the  equation for the gravitino field
(in case of single chiral multiplet $\chi$ and vanishing 
Im~$\Phi$)
$R^{\mu}=m(\tau) \gamma^{\mu\nu} \psi_{\nu}$,
where $ \gamma^{\mu\nu}={1 \over 2}
\left( \gamma^{\mu}\gamma^{\nu}- \gamma^{\nu}
 \gamma^{\mu}\right)$ and 
 gravitino mass $m=m(\phi(\tau))$ is given by
$m = e^{K/2}\, {W\over M_P^2}$. 
This equation can be transformed into the form
\begin{equation}
 \not\!\! {\cal D} \psi _\mu + m \psi _\mu  =
  \left( {\cal D}_\mu -\frac{m}{2}\gamma_\mu\right) \gamma^\nu
  \psi_\nu\, .
   \label{master}
\end{equation}
The gravitino equation (\ref{master}) is a curved spacetime
generalization of the familiar gravitino equation
 $(\not\! {\partial} +m_0) \psi_{\mu}=0$
 in a
flat metric, where $m_0$ is a constant gravitino mass.
The generalization  of the first constraint  equation
$\partial^{\mu}\psi_{\mu}=0$ can be obtained from the equality
$\gamma_{\mu}R^{\mu}=2\gamma^{\mu\nu} {\cal D}_{\mu}\psi_{\nu}$  and
reads
\begin{equation}
{\cal D}^\mu\psi_\mu-\not\!\! {\cal D} \gamma^{\mu} \psi_{\mu} +
{3 \over 2} m \gamma^\mu \psi_\mu =0 \ .\label{constr3}
\end{equation}
The generalization of the
 second  constraint  equation  $\gamma^{\mu} \psi_{\mu}=0$
can be obtained from the equality $ {\cal D}_{\mu}R^{\mu}=
-{1 \over 2}G_{ \nu\rho  }\gamma^\nu \psi^\rho$ (dropping the torsion
term)  and is
\begin{equation}
{3 \over 2} m^2\gamma^\mu \psi_\mu+(\gamma^\mu \partial_{\mu}m)\gamma^\nu \psi_\nu-
(\partial_{\mu}m)\psi^\mu=
-{1 \over 2}G_{\mu \nu }\gamma^\mu \psi^\nu \, . \label{constr1}
\end{equation}
In our case $m=m(\tau)$  and from (\ref{constr1}) one can find an
algebraic relation between $ \gamma^0\psi_0$ and $\gamma^i
\psi_i$:
\begin{equation}
\gamma ^0\psi_{0}=\hat A  \gamma ^i \psi_i   \ . \label{A}
\end{equation}
Here $\hat A$ is a matrix
 which will  play a
 crucial role in our  description of the interaction of the 
gravitino with the varying background fields. If $\rho$ and $p$
are the background energy-density and pressure, we have $G^0_0=
M^{-2}_P\rho$,\, $ G^i_k=-  M^{-2}_P p\, \delta^i_k$, and one can
represent the matrix $\hat A$ as follows:
\begin{equation}
  \hat A= {{ p-3m^2M^2_P}
 \over { \rho+ 3  m^2M^2_P }}\,
+\,\gamma_0{{2 m' a^{-1}M^2_P}
 \over { \rho+ 3  m^2 M^2_P}} =A_1+\gamma _0A_2\,.
\label{A2}
\end{equation}
We shall  solve the equation (\ref{master}) using the constraint equations
in the form (\ref{constr1}) and (\ref{A}). We use a plane-wave
ansatz $\psi_{\mu} \sim e^{i {\bf k \cdot x}}$ for the
space-dependent part. Then $\psi_i$ can be decomposed into
  its  transverse part
$\psi^T_i$, and to the longitudinal part $\psi^L_i$ which is defined by the
 trace $\gamma^i\psi_i$.
 Two degrees of freedom of $\psi_{\mu}$ are
associated with the transverse part $\psi^T_i$, which correspond
to helicity $\pm3/2$ and two degrees of freedom are associated with
 $\gamma^i\psi_i$ (or $\psi_0$)
which correspond to helicity $\pm1/2$.
For the helicity $\pm 3/2$ states we have to derive the equation
for $\psi^T_i$. We apply decomposition $\psi=\psi^T_i+\psi^L_i$ to the master
equation (\ref{master}) for $\mu=i$ and obtain
\begin{equation}
\left(\gamma ^{\mu} 
\partial _\mu + \frac{ a'}{2a}\gamma^0 +
 ma \right)\psi_i^T=0 \,.
\label{trans}
\end{equation}
 The transformation
$\psi^T_i=a^{-1/2}\Psi^T_i$ reduces
 the equation for the transverse part to the
free Dirac equation with a time-varying mass term $ma$, c.f. eq. (\ref{5}).
In the previous section we explained 
 how to treat this type of equation.
The essential part  of $\Psi^T_i$ is given by the time-dependent
part of the eigenmode  of the transversal component $X_T(\eta)$,
which obeys second-order equation (c.f. (\ref{100})):
\begin{equation}
X_T''+\left(k^2+(ma)^2-i(ma)'\right)X_T=0 \ .
 \label{dirac}
\end{equation}
The corresponding equation for gravitino with helicity 1/2 is more
complicated. We have to find $k^i\psi_i$ and $\gamma ^i \psi_i$.
  The equation for
the components $k^i\psi_i$ can be obtained from the constraint
equation (\ref{constr3}).
The  equation for $\gamma ^j\psi _j$
can be derived from (\ref{master}).
Using $\psi_{0}$ from  (\ref{A}),
 we get an
equation for $\gamma^i\psi_i$
\begin{equation}
 \left( \partial _\eta +\hat{B}-
i{\bf k  \cdot  \gamma} \gamma _0 \hat{A} \right)
 \gamma^i \psi_i=0\,,
 \label{trace1}
\end{equation}
where
$ \hat  B =-\frac{3a'}{2a}\hat{A}-{ {m a} \over 2}\gamma _0(1
+3\hat{A}$. 
The time-dependent factor of the spinor $\gamma^i\psi_i$, which we denote as
$f_k(\tau )$, obeys a second-order differential equation.
  By the
substitution $f_k(\tau ) = E(\eta)X_L(\eta)$, with
$E=(-A^*)^{1/2} \exp\left(-
 \int^\eta d\eta\, Re B\right),$
  the equation   for the function $f_k(\tau )$  is reduced to the
final oscillator-like equation for the time-dependent mode
function $X_L(\eta)$:
\begin{equation}
X_L''+\left(\vert A \vert^2k^2+\Omega_L^2-i\Omega_L'\right)X_L=0 \,.
 \label{final}
\end{equation}
Here $
a^{-1} \Omega_L= {i \over 2} \partial_{\tau} \ln A^*+
{{3 a'} \over {2a}}A_2+{1 \over 2}ma \left(-1+3A_1 \right)$.
For an arbitrary background FRW metric $a(\tau)$ and
background scalar field $\phi(\tau)$, equation (\ref{final})
 may lead to ill-defined
physics. For instance, if $\vert A \vert > 1$, it describes noncasual
propagation of spin 3/2 particles. In the context of preheating,
the background  fields  $\phi(\tau)$ and $a(\tau)$
 are oscillating.  Naively, one
 could expect that $\vert A \vert$ is oscillating too. In this case one
 would reach  a pathological conclusion that  the strongest
parametric excitation  of spin 3/2 eigenmodes will be
 with the highest momenta $k$.
Noncausality and other defects of  spin 3/2 fields interacting with 
an external electromagnetic field are well known \cite{VZ}.
It is interesting  to compare the equations for 
a spin 3/2 field in an external  electromagnetic field and 
in  external gravitational/scalar fields. Then we will show
how the apparent defects of  (\ref{final}) are resolved in supergravity.

\subsubsection{Spin 3/2 Field Equations in  an External Electromagnetic Field}
The equations for a charged spin 3/2 field interacting with an external
 electromagnetic field
$A_{\mu}$ with field strength tensor $F_{\mu\nu}$ in flat space-time were
 derived in \cite{VZ}.
After simple manipulations, they can be re-written in the form 
of the equation of motion
\begin{equation}
 \not\!\! \hat {\cal D} \psi _\mu + m_0 \psi _\mu  =
  \left( \hat {\cal D}_\mu -\frac{m_0}{2}\gamma_\mu\right) \gamma^\nu
  \psi_\nu\, ,  \label{master1}
\end{equation}
and two constraint equations
\begin{eqnarray}
&&\hat {\cal D}^\mu\psi_\mu-\not\!\! \hat {\cal D} \gamma^{\mu} \psi_{\mu} +
{3 \over 2} m_0 \gamma^\mu \psi_\mu=0 \ ,\label{constr31}\\
%\end{equation} 
%\begin{equation}
&&{3 \over 2} m_0^2\gamma^\mu \psi_\mu=
-{1\over 2 } {\tilde F}^{\mu\nu}\gamma_{\mu}\psi_{\nu}\, , \label{constr11}
\end{eqnarray}
where ${\tilde F }^{\mu\nu}=e \gamma^5  \epsilon^{\mu\nu\rho\sigma}
F_{\rho\sigma}$. 
Here $\hat {\cal D}_{\mu}\equiv \partial_{\mu}-ieA_{\mu}$.
The system of equations (\ref{master1}), (\ref{constr31}), (\ref{constr11})
for flat case with EM background
is similar to the  equations (\ref{master}), (\ref{constr3}), (\ref{constr1})
for cosmological problem
up to definition of derivatives  $\hat {\cal D}_{\mu}$ instead
of  ${\cal D}_{\mu}$, a
constant  mass  $m_0$ instead of the altering mass $m(\tau)$, and 
$ {\tilde F}^{\mu\nu}$ instead 
of  $ G^{\mu\nu}$. 
It is interesting that a spin 3/2 field interacting with external fields
can be described in the unified way.
In \cite{VZ} it was shown that for some configurations 
of the electromagnetic field, the 
propagation of spin 3/2 field violates causality.

\subsubsection{Self-consistent Gravitino Problem in Supergravity}
Contrary to the inconsistent setting of the problem of 
a spin 3/2 field in an arbitrary EM or gravitational field,
the spin 3/2 gravitino field in supergravity  should  be
consistent. In this section we will show that the gravitino equations
in an expanding universe with moving scalars are consistent.
We will use a  model with  a single chiral multiplet.
Let us concentrate on the matrix $\hat A$ given by (\ref{A2}).
 In the models
where the energy-momentum tensor is determined by the energy of a
classical scalar field and $\Phi$ depends only on time we have
$\rho = |\dot \Phi|^2 +V$,
 $ p = |\dot \Phi|^2 -V$. The scalar potential is
$V(\Phi) = e^K\, \left|{\cal D} W\right|^2-{ 3 m^2 M_P^2}$.
Also, we have $m'=a e^{K/2}{\cal D} W \dot \Phi/M_P^2$.
Therefore, the matrix $\hat A$ can be rewritten in terms of $\dot \Phi$
and $e^{K/2}{\cal D} W$ only
\begin{equation}
  \hat A= {{ |\dot \Phi|^2-  \left|e^{K/2}  {\cal D} W\right|^2 }
 \over {  |\dot \Phi|^2+  \left| e^{K/2}   {\cal D} W\right|^2   }}\,
+\,\gamma_0{{2  \dot \Phi   e^{K/2}{\cal D} W     }
 \over { |\dot \Phi|^2+ \left| e^{K/2}{\cal D} W\right|^2 }}\,.
\label{A3}
\end{equation}
From this form of $\hat A$ it follows that
\begin{equation}
\vert A \vert^2 \equiv A_1^2+A_2^2=1  \label{A4}
\end{equation}
for an arbitrary superpotential $W$.
Thus 
 $ A$  can be represented as
 $ A =-\exp \left(2{ i}\int_{-\infty }^t dt \,\mu (\eta)\right)$.
Using the Einstein equations, one obtains 
$ \mu ={\cal D}{\cal D}W + \Delta$,
where the correction  $\Delta={\cal O}(M_P^{-1})$ is given in \cite{grav}.
The expression for $\mu$ becomes much simpler and its 
interpretation is more transparent if the amplitude of
oscillations of the field $\Phi$ is much smaller than $M_P$. In
the limit $\Phi/M_P \to 0$ one has
  $\mu = \partial_\Phi \partial_\Phi W $.
This coincides  with the mass of both fields of the chiral
multiplet (the scalar field and spin 1/2 fermion) in rigid
supersymmetry. When supersymmetry is spontaneously broken, the
chiral fermion, goldstino, is `eaten' by the gravitino which becomes
massive and acquires helicity $\pm 1/2$ states in addition to
the helicity $\pm 3/2$ states of the massless gravitino.
With this form of $\hat A$ the gravitino equation 
(\ref{final})  becomes 
\begin{equation}
X_L''+\left( k^2+\Omega_L^2-i\Omega_L'\right)X_L=0 \,.
 \label{final11}
\end{equation}
with
$a^{-1} \Omega_L=
\mu-\frac{3}{2} H\sin 2{\int \mu dt }- \frac{1}{2}
m \left(1 +3 \cos{2\int \mu dt }\right)$.
Equation (\ref{final11}) is consistent and describes
the creation of gravitinos from preheating.
The solution of the consistency problem in the equation 
(\ref{final11})  is that the symmetries imprinted to
the Supergravity provide $\vert A \vert=1$. But it does not
mean that $A$ is a constant.
The matrix $\hat A$ does not become constant even  in the limit $M_P \to
\infty$. The phase of $\hat A$ rotates when the background
scalar field oscillates. The amplitude and sign of $A$ change two
times within each oscillation. Consequently, the relation between
$\gamma ^0\psi_{0}$ and $ \gamma ^i \psi_i$ also oscillates during
the field oscillations. This means that the gravitino with
helicity 1/2 (which is related to $\psi_0$) remains coupled to the
changing background even in the limit $M_P \to \infty$. In a
sense, the gravitino with helicity 1/2 remembers its goldstino
nature. This is the main reason why gravitino production in
this background in general is not suppressed by the gravitational
coupling. The main dynamical quantity which is responsible for the
gravitino production in this scenario will not be the small
changing gravitino mass $m(t)$, but the mass of the chiral
multiplet $\mu$, which is much larger than $m$.

\subsection*{Problem of  Gravitino Over-Production from Preheating}
 As an example, consider the model with the
superpotential $W= \sqrt{\lambda}\Phi^3/3$.
The parameter $\mu$  for this model is given by
$\mu=\sqrt{2\lambda} \phi$. It rapidly changes in the interval
between $0$ and $\sqrt{2\lambda} \phi_0$. Initially it is of the
same order as $H$ 
and $m$, but then $H$ and $m$ rapidly decrease
as compared to $\mu$, and therefore the oscillations of $\mu$
remain the main source of the gravitino production. In this case
the production of gravitinos with helicity 1/2 is much more efficient
than that of helicity 3/2.
The theory of production of gravitinos with helicity 1/2 in this
model is similar to the theory of production of spin 1/2 fermions
with mass $ \sqrt{2\lambda}\phi$ by the coherently oscillating
scalar field in the theory $\lambda\phi^4/4$, which we considered
above.   Growth of  helicity 1/2 gravitino
 modes (\ref{final11}) occurs
in the non-perturbative regime of parametric excitation. The modes
get fully excited with occupation numbers $n_k \simeq 1$
within
about ten oscillations of the field $\phi$, and the width of the
parametric excitation of fermions in momentum space is about
$\sqrt\lambda\phi_0$. 
This result violates the cosmological constraints on the abundance
of gravitinos with mass $\sim 10^2$ GeV by 4 orders of magnitude
\cite{grav}. The most dangerous gravitino over-production (by 14 orders
of magnitude) occurs in the class of inflationary models
where $V(\phi)$ does not have a minimum and where
preheating is gravitational (NO models) \cite{FKL1}.
Thus the investigation of the non-thermal gravitino production in
the early universe may serve as a useful tool helping us to
discriminate among various versions of the cosmological theory.


\begin{references}
\bibitem{SUGRA} E. Cremmer, S. Ferrara, L. Girardello and A. Van Proeyen,
{\it Nucl.\ Phys.} {\bf B212} (1983).
\bibitem{FKL1} G. Felder, L.A. Kofman, and A.D. Linde, hep-ph/9903350 and
{\it Phys.\ Rev.} {\bf D 59}, 123523 (1999), hep-ph/9812289.
\bibitem{grav} R. Kallosh,  L. Kofman,   A.D. Linde, and A. Van
Proeyen, hep-th/9907124
\bibitem{GK} P. Greene  and L. Kofman, {\it Phys.\ Lett.} {\bf 448},  6 (1999).
\bibitem{KLS}  L. Kofman, A. Linde and  A. Starobinsky,  {\it Phys. Rev. Lett.}
{\bf 73}, 3195 (1994), hep-th/9405187; L. Kofman, A. Linde and A.
Starobinsky, {\it Phys. Rev.} {\bf D56},  3258 (1997), hep-ph/9704452;
\bibitem{GKLS}P. B. Greene, L. Kofman, A.D. Linde, and A.A. Starobinsky, 
{\it Phys.\ Rev.} D {\bf 56}, 6175-6192 (1997), hep-ph/9705347.
\bibitem{VZ} G. Velo and D. Zwanziger, 
{\it Phys.\ Rev.} {\bf 186}, 1337 (1969)
\end{references}
\end{document}